\documentclass[%
reprint,
twocolumn,
nobalancelastpage,
superscriptaddress,
 amsmath,amssymb,
 aps,
rc]{revtex4-2}
\usepackage{siunitx}
\usepackage{hyperref}
\usepackage{float}
\usepackage{amsmath,amsthm,amsfonts,amssymb,amscd,mathrsfs,mathtools,physics,esint,cancel,bm}

\begin{document}

\preprint{APS/123-QED}

\title{Multiplexed color centers in a silicon photonic cavity array
}

\author{Lukasz Komza}
\affiliation{Department of Physics, University of California, Berkeley, Berkeley, California 94720, USA}
\affiliation{Materials Sciences Division, Lawrence Berkeley National Laboratory, Berkeley, California 94720, USA}

\author{Xueyue Zhang}
\affiliation{Department of Electrical Engineering and Computer Sciences, University of California, Berkeley, Berkeley, California 94720, USA}
\affiliation{Department of Physics, University of California, Berkeley, Berkeley, California 94720, USA}

\author{Hanbin Song}
\affiliation{Materials Sciences Division, Lawrence Berkeley National Laboratory, Berkeley, California 94720, USA}
\affiliation{
Department of Materials Science and Engineering, University of California, Berkeley, Berkeley, California 94720, USA
}

\author{Yu-Lung Tang}
\affiliation{Department of Physics, University of California, Berkeley, Berkeley, California 94720, USA}
\affiliation{Materials Sciences Division, Lawrence Berkeley National Laboratory, Berkeley, California 94720, USA}

\author{Xin Wei}
\affiliation{Department of Electrical Engineering and Computer Sciences, University of California, Berkeley, Berkeley, California 94720, USA}

\author{Alp Sipahigil}
\email{Corresponding author: alp@berkeley.edu}
\affiliation{Department of Electrical Engineering and Computer Sciences, University of California, Berkeley, Berkeley, California 94720, USA}
\affiliation{Materials Sciences Division, Lawrence Berkeley National Laboratory, Berkeley, California 94720, USA}
\affiliation{Department of Physics, University of California, Berkeley, Berkeley, California 94720, USA}

\date{\today}

\begin{abstract}

Entanglement distribution is central to the modular scaling of quantum processors and establishing quantum networks.
Color centers with telecom-band transitions and long spin coherence times are suitable candidates for long-distance entanglement distribution.
However, high-bandwidth memory-enhanced quantum communication is limited by high-yield, scalable creation of efficient spin-photon interfaces. 
Here, we develop a silicon photonics platform consisting of arrays of bus-coupled cavities. The coupling to a common bus waveguide enables simultaneous access to individually addressable cavity-enhanced T center arrays.
We demonstrate frequency-multiplexed operation of two T centers in separate photonic crystal cavities. 
In addition, we investigate the cavity enhancement of a T center through hybridized modes formed between physically distant cavities. 
Our results show that bus-coupled arrays of cavity-enhanced color centers could enable efficient on-chip and long-distance entanglement distribution. 
\end{abstract}

\maketitle
\begin{figure*}[t]
    \centering
    \includegraphics[width=2\columnwidth]{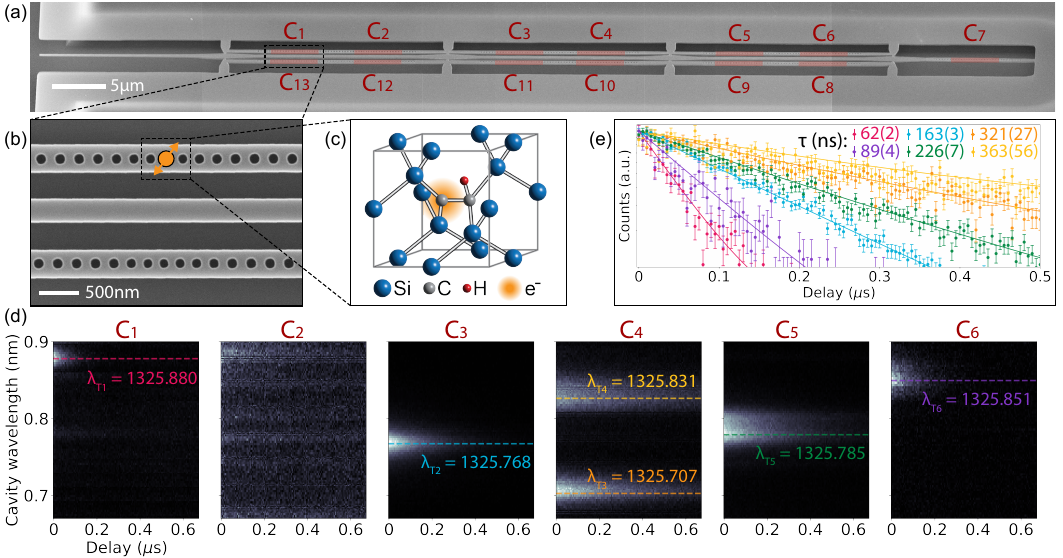}
    \caption{\textbf{T centers in bus-coupled photonic crystal cavities.}
    (a) SEM image of fabricated device. 13 cavities are evanescently coupled to a silicon beam waveguide, which is terminated with a cavity ($C_7$) that acts as a mirror off-resonance.
    (b) Zoomed-in SEM image of bus-coupled photonic crystal cavities.
    (c) Structure of the T center in silicon.
    (d) Time-resolved photoluminescence excitation measurements taken during step-wise tuning of six cavities ($C_1 \to C_6$).
    (e) Fluorescence lifetime ($\tau$) measurements and fits for six cavity-enhanced T centers ($T_1 \to T_6$) in the cavities.}
    \label{fig1}
\end{figure*}

A central challenge in scaling quantum systems lies in the large-scale distribution of entanglement between individual quantum systems. While generating entanglement between local qubits is the basis of quantum computing, extending entanglement beyond local qubits to physically separated quantum systems enables quantum communication \cite{Muralidharan2016, Kimble2008, Wehner2018} and the modular scaling of quantum processors \cite{Monroe2014}. 
Quantum systems with optical interfaces, such as trapped ions~\cite{Moehring2007, Stephenson2020, Krutyanskiy2023}, neutral atoms~\cite{Ritter2012, Hofmann2012}, and color centers~\cite{Bernien2013, Sipahigil2016, Evans2018, Larocque2024}, are natural building blocks for quantum networks due to the low propagation loss of photons over long distances. In particular, color centers in silicon have emerged as promising candidates due to their telecom-band photon emission and compatibility with silicon device fabrication~\cite{Redjem2020, Durand2021, Islam2023, Johnston2024, Larocque2024, Komza2024, PhotonicInc2024}. These properties make them well-suited for integration in silicon photonics, a mature platform offering a toolbox of active and passive components~\cite{Sun2015, ErrandoHerranz2020}. 

Among these color centers, T centers feature long spin lifetimes and photon emission in the telecom O-band, making them an attractive platform for realizing quantum repeater nodes~\cite{Bergeron2020, Higginbottom2022, DeAbreu2023}. However, their long optical lifetimes ($\sim1\,\mathrm{\mu s}$) make realizing an efficient optical interface challenging. While recent work has demonstrated the enhancement of photon emission rates through integration in photonic crystal cavities~\cite{Islam2023, Johnston2024}, achieving a high and consistent yield of bright cavity-enhanced T centers in a single device remains a challenge. This is due to the relatively low creation yield of T centers \cite{MacQuarrie2021} combined with the small mode volumes of photonic crystal cavities. Accessing multiple cavities at once by coupling cavities to a common bus waveguide could improve device yield and enable wavelength-division multiplexing of color centers. This would improve entanglement generation rates limited by photon propagation times and open the door to the parallel operation of color centers in the spectral domain. Furthermore, bus-mediated interactions between cavities and emitters could enable scalable on-chip entanglement generation between silicon color centers by utilizing the inherently shared spatial mode~\cite{Kobayashi2016}.

In this work, we develop a photonics platform based on arrays of waveguide-coupled photonic crystal cavities containing T centers in silicon. We demonstrate multiplexed operation of two T centers in separate cavities through a bus waveguide. Furthermore, we demonstrate the hybridization of spatially separated cavities through bus-mediated interactions and study the Purcell enhancement of a single T center through delocalized cavity modes.

\noindent{}\textbf{Bus-coupled photonic crystal cavity arrays.}
Our device consists of arrays of 1D photonic crystal cavities evanescently coupled to a common single-mode bus waveguide (Fig~\Ref{fig1}(a,b)). This device geometry enables simultaneous access to an array of cavity resonances through the bus waveguide. The bus waveguide is adiabatically tapered on one end to mode match with a lensed fiber with measured coupling efficiencies up to $40$\%. On the other end, the bus waveguide is terminated with a cavity, acting as a mirror for non-resonant wavelengths and enabling single-sided operation of the device. We create T centers in the device through ion implantation and rapid thermal annealing (Fig~\Ref{fig1}(c)). Device design and fabrication details are available in Appendix~\ref{fab-and-design}.

\noindent{}\textbf{T centers in cavity arrays.} We systematically identify single T centers in cavities by performing time-resolved photoluminescence excitation (PLE) measurements during step-wise tuning of individual cavity resonances through the T center inhomogeneous distribution (Fig~\Ref{fig1}(d)). The tuning mechanism is based on the deposition and site-selective removal of thin nitrogen films, which we describe in further detail in the next section. Single strongly-enhanced T centers are clearly visible as an increased fluorescence signal at specific cavity wavelengths. We obtain cavity-enhanced lifetimes by fitting the exponentially decaying fluorescence (Fig~\Ref{fig1}(e)). The data in Fig~\Ref{fig1}(d,e) is a representative subset of the data from six cavities ($C_1 \to C_6$) where six T centers ($T_1 \to T_6$) are identified. The Purcell-enhanced lifetimes range from $62(2)\,\mathrm{ns}$ to $363(56)\,\mathrm{ns}$, reduced from the natural lifetime of $\sim940\,\mathrm{ns}$~\cite{Bergeron2020}. For $T_1$ ($\lambda_{T1}=1325.880\,\mathrm{nm}$), we measure a lifetime of $62(2)\,\mathrm{ns}$, corresponding to a lower-bounded Purcell factor of $61$ (Appendix~\ref{purcell-analysis}). We calculate a minimum emitter-cavity coupling $g/2\pi=115\,\mathrm{MHz}$ from the Purcell enhancement, compared to the maximum coupling $g/2\pi=400\,\mathrm{MHz}$ expected from cavity mode volume simulations ($V = 0.5(\lambda_0/n)^3$). The discrepancy between the calculated and maximum coupling is due to uncertainty in the T center dipole orientations and the relative positions of the T centers relative to the cavity modes. We fit the power-broadened emitter linewidth ($\gamma/2\pi = 3.0\,\mathrm{GHz}$) and cavity linewidth ($\kappa/2\pi = 5.1\,\mathrm{GHz}$) to calculate the lower bound on cooperativity $C = 4g^2/(\kappa \gamma) =  3.5\times10^{-3}$. Further characterization of other T centers and cavities can be found in Appendix~\ref{maps}.

\begin{figure}[t!]
    \centering
    \includegraphics[width=\columnwidth]{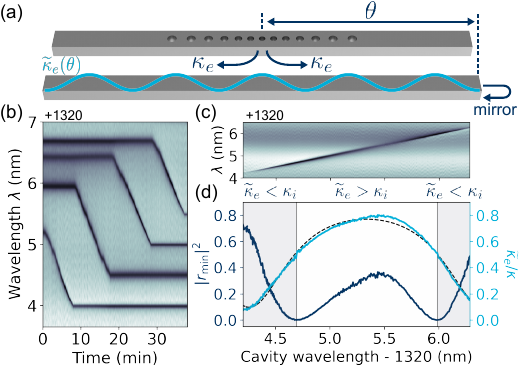}
    \caption{\textbf{Site-selective cavity resonance tuning.} 
    (a) Schematic showing a cavity externally decaying to left- and right-propagating modes at rate $\kappa_e$. Interference of the two paths results in a periodic modulation of the effective external cavity decay $\widetilde{\kappa}_e$ as a function of the propagation phase between the cavity and mirror $\theta$.
    (b) Automated generation of a uniformly-spaced array of cavity resonances using resonant tuning through the bus waveguide.
    (c) Tuning of a single cavity over $2\,\mathrm{nm}$ results in periodic modulation of $\widetilde{\kappa}_e$.
    (d) Extraction of minimum reflectivities from (c) (left). Minimum values determine the external coupling fraction $\widetilde{\kappa}_e/\kappa$ (right). The black dashed line is the fit of $\widetilde{\kappa}_e/\kappa$.}
    \label{fig2}
\end{figure}

\noindent{}\textbf{Programmable cavity tuning.}
Programmable and selective tuning of individual cavities is critical in our multi-cavity platform for aligning cavities to T centers and facilitating interactions between distinct cavities. We achieve this by first red-shifting all cavity resonances through the global deposition of a thin film of nitrogen on the device, increasing the effective refractive index~\cite{Mosor2005, Strauf2006}. We subsequently remove the film in a cavity-selective way by resonantly exciting a target cavity through the bus waveguide. The resonant build up of the intra-cavity field results in higher intensities in the target cavity compared to off-resonant cavities and the bus waveguide. Above a power threshold, absorptive heating in the cavity causes the local nitrogen ice to sublimate and blue-shift the target cavity's resonance. We implement a closed-loop resonant tuning method where we adjust the laser power while monitoring cavity resonance positions to maintain a desired tuning rate (Appendix~\ref{experimental-setup}). We use this method to programmatically generate evenly-spaced cavity resonances using resonant pulses applied to the bus waveguide in Fig~\Ref{fig2}(b). This method fails when cavities are not spectrally resolvable or when two cavities overlap spectrally. In these cases, we focus an above-bandgap laser through an objective on a target cavity location to locally heat it and blue-shift its resonance.

As we tune cavities over a wide wavelength range, we observe periodic modulations in cavity linewidths and reflection depths  (Fig.~\ref{fig2}(c)). The bi-directional coupling of a cavity to the bus waveguide results in interference between the left- and right-propagating emission after reflection at the terminating mirror. We fit the periodic modulation of $\kappa$ due to interference using $\kappa = \widetilde{\kappa}_e + \kappa_i = \kappa_{e} (1 + \cos(2\theta)) + \kappa_i$, where $\kappa_e$ is the one-directional external coupling to the bus waveguide, $\kappa_i$ is the intrinsic decay rate, and $\theta$ is the phase length between the cavity and the terminating mirror (Fig.~\ref{fig2}(a)). The phase length $\theta$ depends on the cavity position and wavelength, resulting in the cavity linewidth modulation as a function of wavelength. By extracting the minimum value in the reflection spectrum $|r_\mathrm{min}|^2 = (2\widetilde{\kappa}_e/(\widetilde{\kappa}_e + \kappa_i)-1)^2$, we can determine the external coupling fraction $\widetilde{\kappa}_e / (\widetilde{\kappa}_e + \kappa_i) = \widetilde{\kappa}_e / \kappa$, plotted in Fig~\Ref{fig2}(d). The cavity is initially strongly under-coupled ($\widetilde{\kappa}_e / \kappa \sim 0$) and barely visible in the reflection spectrum due to destructive interference. As the cavity is blue-shifted, $\widetilde{\kappa}_e$ increases, becoming critically coupled ($\widetilde{\kappa}_e / \kappa = 0.5$) before reaching a maximum value ($\widetilde{\kappa}_e / \kappa = 0.80$) for constructive interference. At this maximum, $\kappa_e/2\pi = 22.2\,\mathrm{GHz}$ and $\kappa_i/2\pi = 5.5\,\mathrm{GHz}$, corresponding to the strongest external coupling determined by the device geometry.

\begin{figure}[t!]
 	\centering
  	\includegraphics[width=\columnwidth]{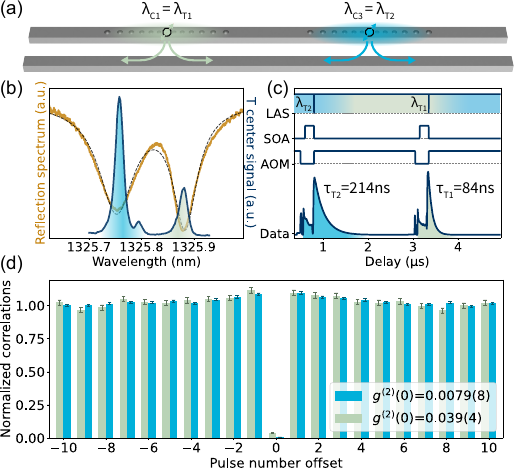}
 	\caption{\textbf{Multiplexed single-photon emission.} 
  (a) Schematic illustrating two simultaneously-enhanced and spectrally- and spatially- distinct T centers.
  (b) PLE spectrum overlaid with the cavity reflection spectrum (orange). The dashed line is the fit of the reflection spectrum ($\kappa_{C1}/2\pi = 9.6\,\mathrm{GHz}$, $\kappa_{C3}/2\pi = 14.2\,\mathrm{GHz}$). (c) Histogram showing time-resolved fluorescence from two cavity-enhanced T centers. AOM and SOA pulse sequences are synchronized with the laser wavelength switching (LAS). (d) Multiplexed $g^{(2)}$ functions for both T centers calculated from correlations in (c).}
 \label{fig3}
\end{figure}

\noindent{}\textbf{Multiplexed cavity enhancement of two T centers.}
By leveraging the programmable tuning of our cavity arrays, we demonstrate parallel operation of two spatially and spectrally separated cavity-enhanced T centers through a single waveguide (Fig.~\ref{fig3}(a)). We choose T centers ($T_1$, $T_2$) where their spectral separation ($\lambda_{T1}=1325.880\,\mathrm{nm}$, $\lambda_{T2}=1325.768\,\mathrm{nm}$, $\Delta_{T1,T2}/2\pi = 19\,\mathrm{GHz}$) is larger than the cavity linewidths ($\kappa_{C1}/2\pi = 9.6\,\mathrm{GHz}$, $\kappa_{C3}/2\pi = 14.2\,\mathrm{GHz}$ at $\lambda_{T1}$, $\lambda_{T2}$) to avoid hybridization between the cavity modes. We align the cavities to their T centers (Fig.~\ref{fig3}(b)) and measure Purcell-enhanced lifetimes of $\tau_{T1} = 84.32(12)\,\mathrm{ns}$ and $\tau_{T2} = 213.89(17)\,\mathrm{ns}$ (Fig.~\ref{fig3}(c)).


\begin{figure*}[t!]
    \centering
    \includegraphics[width=2\columnwidth]{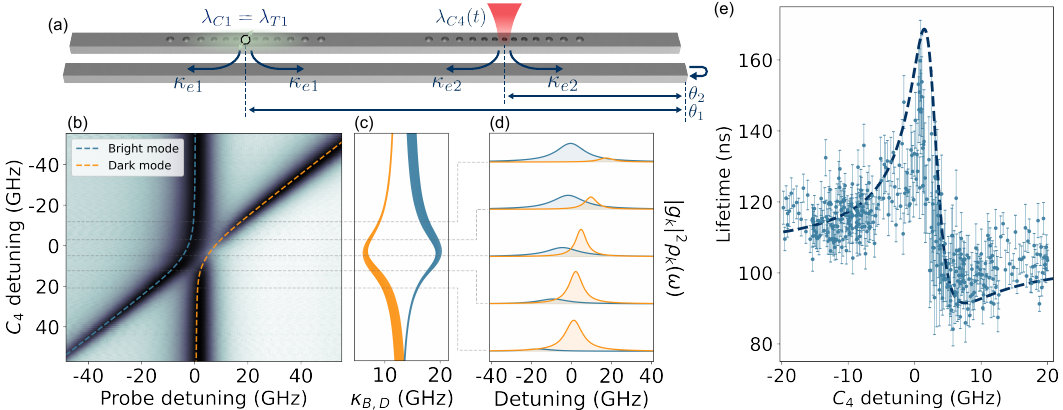}
    \caption{\textbf{Emitter enhancement through a hybridized mode. } 
    (a) Schematic illustrating cavity hybridization. The cavities $C_1$, $C_4$ are separated by phase lengths $\theta_1$, $\theta_2$ from the mirror. $C_1$ is initially resonant with T center $T_1$ ($\lambda_{C1} = \lambda_{T1}$), while $C_4$'s resonance $\lambda_{C4}(t)$ is tuned with a free-space above-gap laser.
    (b) Device reflection spectra during two-cavity hybridization showing the formation of a bright (blue) and dark (orange) hybrid mode. Horizontal dashed lines indicate the slices of the data used in (d).
    (c) Hybrid mode decay rates. (d) Local density of states in $C_1$ from both hybrid modes. The density of states is weighted by the populations of the modes in $C_1$.
    (e) Lifetime of a T center in cavity $C_1$ during hybridization with $C_4$. The dashed line is the lifetime predicted from the model.
    The detunings in (b,d,e) are defined relative to the T center ($\lambda_{T1} = 1325.880\,\mathrm{nm}$).}
    \label{fig4}
\end{figure*}

To multiplex single photons from these T centers, we use a fast tunable laser capable of switching between arbitrary wavelengths in a $35\,\mathrm{nm}$ range in less than $2\,\mathrm{\mu s}$. We use a semiconductor optical amplifier to create $200\,\mathrm{ns}$-long optical pulses with $1\,\mathrm{ns}$ rise and fall times to resonantly excite the T centers. To avoid heating and saturation effects of the SNSPDs, we optically gate the collection path with an acousto-optic modulator. By synchronizing the laser's wavelength switching with the pulse generation and detection, we create a time-multiplexed signal from the single photon emission of two distinct T centers (Fig.~\ref{fig3}(c). To verify the single-photon nature of the signal, we measure intensity correlations in the individual T center emission windows (Fig.~\ref{fig3}(d)) and confirm high quality single-photon emission from both T centers ($g^{(2)}_{T1}(0) = 0.039(4)$, $g^{(2)}_{T2}(0) = 0.0079(8)$). The number of T centers which could be operated in parallel is limited by the ratio of the T center inhomogeneous distribution to the cavity linewidths in the device. When the detuning between two cavities is small compared to their linewidths, bus-mediated interactions result in strong hybridization of cavity modes. We explore the rich physics of this regime in the next section.

\noindent{}\textbf{Hybrid mode enhancement of a T center.} 
The combination of the high degrees of connectivity and tunability in our device presents a platform to explore bus-mediated interactions in integrated quantum photonics. We study these interactions by enhancing a single T center with a delocalized mode formed through the bus-mediated hybridization of two photonic crystal cavities. The cavity mode hybridization can be described by coherent and dissipative interactions, which can be expressed through the effective interaction Hamiltonian~\cite{Lalumire2013}
\begin{equation}
    \hat{H}_{\mathrm{eff}} = \hbar \sum_{l<m} \left( g_{lm} -i\frac{\kappa_{c,lm}}{2} \right) \left( \hat{a}^\dagger_l \hat{a}_m + \hat{a}^\dagger_m \hat{a}_l \right)
\end{equation}
where $l,m$ are the cavity indices, $g_{lm}$ is the coherent coupling, $\kappa_{c,lm}$ is the correlated decay, and $\hat{a}_{l,m}$ ($\hat{a}^\dagger_{l,m}$) is the annihilation (creation) operator for the cavity modes. The coherent interaction is mediated by dispersive coupling to a continuum of photonic modes in the bus waveguide, and the correlated decay is caused by interference of the emission from the cavities. Both interactions depend on the phase lengths between the cavities and the mirror $\theta_{l,m}$. In the absence of a mirror, these bus-mediated interactions depend on the relative phase between the cavities $\theta_l - \theta_m$. In this case, strong coupling cannot be achieved for any phase length condition, giving $4 g_{lm}^2 / (\kappa_l \kappa_m ) \leq 1$~\cite{Lalumire2013}. The addition of a mirror introduces the additional interference path with phase length $\theta_l + \theta_m$. The addition of this path enables access to the strong coupling regime, and we observe signatures of strong coupling such as avoided crossings (Appendix~\ref{SLH-strong}).

We experimentally probe these interactions by tuning cavity $C_4$ across the resonance of a stationary cavity $C_1$ using the above-bandgap laser. To quantitatively understand the cavity hybridization, we build an analytical model that allows us to extract key parameters (Appendix~\ref{SLH-model}). We measure cavity linewidths prior to hybridization of $\kappa_{c_1}/2\pi = 13.8\,\mathrm{GHz}$ and $\kappa_{c_4}/2\pi = 12.4\,\mathrm{GHz}$. The cavity modes hybridize as $C_4$ approaches $C_1$, where one of the hybridized modes appears darker and the other mode appears brighter in the reflection spectrum (Fig.~\ref{fig4}(b)). This change in contrast of the dark (bright) mode is a result of reduced (enhanced) external coupling rates from destructive (constructive) interference in the correlated dissipation (Fig.~\ref{fig4}(c). The different interference behavior of the dark and bright modes is due to the relative phases between the two cavities in the hybrid modes. At maximum hybridization, the decay rates are measured to be $\kappa_{B}/2\pi = 19.6\,\mathrm{GHz}$ for the bright mode and $\kappa_{D}/2\pi = 6.6\,\mathrm{GHz}$ for the dark mode. The correlated dissipation is calculated from the model to be $\kappa_c/2\pi = 6.6\,\mathrm{GHz}$. While the correlated dissipation results in relative changes of the external coupling rates, the coherent interaction leads to detuning between the hybridized modes when the two cavities are on resonance. The strength of the coherent interaction calculated from the model is $g_c/2\pi = 4.6\,\mathrm{GHz}$. 

We further study these interactions by analyzing the Purcell enhancement of a T center through delocalized hybrid modes. Cavity $C_1$ is initially tuned to be near-resonant with one of its T centers ($\lambda_{T1} = 1325.880\,\mathrm{nm}$). We monitor the T center's lifetime during cavity hybridization in Fig.~\ref{fig4}(e). In the Purcell regime, the optical decay rate can be calculated through Fermi's golden rule as $\Gamma \propto \sum_k \left| g_k\right|^2 \rho_k(\omega)$, where $g_k$ are coupling constants to cavity modes and $\rho_k(\omega)$ is the photonic density of states. The coupling terms $g_k$ are proportional to the amplitudes of the hybrid modes in $C_1$, while the maximum density of states of mode $k$ is proportional to the inverse of its linewidth. By plotting the density of states $\rho_k(\omega)$ of the hybrid modes weighted by their populations in $C_1$ (Fig.~\ref{fig4}(d)), we visualize the contributions of the hybrid modes to the overall Purcell enhancement of the T center. At the beginning of hybridization, the T center is enhanced primarily by the bright mode, and the dark mode is far detuned. As $C_4$ approaches $C_1$, both hybrid modes are nearly resonant with the T center. However, both modes become slightly detuned from the T center through the coherent interaction, resulting in a lower effective density of states and a longer lifetime. As $C_4$ is tuned past $C_1$, the dark mode becomes resonant with the T center while the bright mode becomes detuned. The dark mode has a smaller decay rate than the bright mode, resulting in a shorter final lifetime. The lifetime predicted from the local density of states by the parameters extracted from the analytical model agrees with the trend of measured lifetime (Fig.~\ref{fig4}(e)). The discrepancy is due to the difficulty of fitting the resonance position of $C_4$ from the reflection spectrum, as there is high uncertainty in the exact position during hybridization due to the formation of a dark mode. This fitting uncertainty manifests as uncertainty in the scaling of the x-axis of Fig.~\ref{fig4}(e).

\noindent{}\textbf{Discussion and outlook.} 
Our results demonstrate multiplexed single photon generation from cavity-enhanced T centers and the coupling of a single T center to a delocalized cavity mode. We discuss the limitations of our platform and consider necessary and realistic improvements to advance this platform for scalable entanglement distribution.

The rate of entanglement distribution over long distances is limited by photon propagation times through optical fibers ($\sim 50\,\mathrm{\mu s}$ for a $10~\mathrm{km}$ link). Commonly used heralded protocols~\cite{Pompili2021, Knaut2024} further multiply these delays by the number of trials needed to successfully generate entanglement. Simultaneous operation of many cavity-enhanced color centers could increase entanglement generation bandwidth via wavelength division multiplexing. The number of T centers which can be multiplexed per bus waveguide is limited by spectral crowding. In the current device, we measure intrinsic cavity quality factors of around $35\times 10^3$, corresponding to $\kappa_i/2\pi \sim 6\,\mathrm{GHz}$. T center transition frequencies in our device are distributed over a $\sim 30\,\mathrm{GHz}$ range, making the parallel operation of more than a few centers challenging due to uncontrolled hybridization between cavity modes. With realistic improvements of cavity linewidths to sub-GHz levels~\cite{DeAbreu2023, PhotonicInc2024}, tens of T centers could be operated in parallel per waveguide. Further spectral multiplexing could be achieved by strain tuning the emitters up to $150\,\mathrm{GHz}$ using micro-electromechanical actuators~\cite{Meesala2018}. Spatial multiplexing is also possible using waveguide arrays and cryogenic fiber array couplers~\cite{Yin2021}. To control individual spins in the T center array platform for spin-photon entanglement, the introduction of magnetic field gradients could make the electronic transitions individually addressable in frequency for multiplexed operation. 

This platform could also be used for the scalable generation of entanglement between color centers on-chip. The inherent sharing of spatial modes between separate T centers in this platform enables direct interference in the frequency domain~\cite{Kobayashi2016}. By leveraging silicon photonics components such as switches~\cite{Lee2014, Seok2016}, single-photon detectors~\cite{Najafi2015, Gyger2021}, and modulators~\cite{Reed2010, Hu2021}, on-chip entanglement generation could be achieved with high efficiency.

The quality of individual nodes is critical to achieving high success rates in heralded entanglement generation. The quality of the optical interface is characterized by the cooperativity $C = 4g^2 / (\kappa \gamma)$, where $g$ is the emitter-cavity coupling, $\kappa$ is the cavity linewidth, and $\gamma$ is the emitter linewidth. We estimate cooperativities ranging from $0.5 \times 10^{-3}$ to $3.5 \times 10^{-3}$ in our device, limited by broad cavity linewidths and emitter linewidths including power broadening. Several approaches can be taken in improving the cooperativity. Cavity linewidths can be reduced through lower implantation densities and optimized fabrication, and emitter linewidths can be decreased with environment engineering through active electronics~\cite{Anderson2019, Day2024}. With realistic improvements of $\kappa/2\pi = 500\,\mathrm{MHz}$ and $\gamma/2\pi = 5\,\mathrm{MHz}$~\cite{DeAbreu2023, PhotonicInc2024}, cooperativities exceeding $10$ could be achieved. With sufficiently high cooperativities $C\sim100$, on-chip entanglement between emitters could be achieved by leveraging the strong and tunable interactions between cavities in this platform~\cite{Grinkemeyer2024}. The addition of photonic phase shifters can dynamically control the cavity hybridization to tailor long-range coherent and dissipative interactions~\cite{Plenio1999, Shah2024}.

\noindent{}\textbf{Acknowledgments.} 
We thank Yiyang Zhi, Zihuai Zhang, Xudong Li, and Niccolo Fiaschi for technical assistance. This work was supported by the Office of Advanced Scientific Computing Research (ASCR), Office of Science, U.S. Department of Energy, under Contract No. DE-AC02-05CH11231 and Berkeley Lab FWP FP00013429. L.K. and A.S acknowledge support from the NSF (QLCI program through grant number OMA-2016245, and  Award No. 2137645). X.Z. acknowledges support from the Miller Institute
for Basic Research in Science. The devices used in this work were fabricated at the Berkeley Marvell NanoLab.

\bibliography{references}

\appendix
\renewcommand{\thefigure}{S\arabic{figure}}
\setcounter{figure}{0}
\section{Experimental setup}\label{experimental-setup}

\begin{figure*}[t!]
 	\centering
 	\includegraphics[width=2\columnwidth]{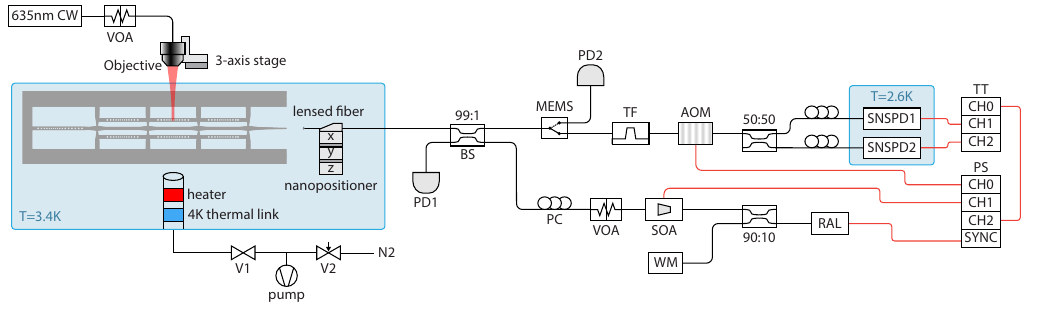}
 	\caption{\textbf{Experimental setup details.} Black and orange lines represent optical fiber and electrical connections respectively. VOA: Variable optical attenuator. PD1: Photodiode used for measuring input power. PD2: Fast photodiode used for taking device reflection spectra. MEMS: Fiber MEMS switch. PC: Polarization controller. SOA: Semiconductor optical amplifier. TF: Tunable filter. AOM: Acousto-optic modulator. WM: Wavemeter. RAL: Random-access laser. SNSPD: Superconducting nanowire single photon detector. TT: Time tagger. PS: Pulse streamer. V1: Solenoid valve. V2: Needle valve.}
\label{figS1}
\end{figure*}

The sample is mounted in a cryostat (Montana Instruments Cryostation s200) and cooled down to 3.4 K. A lensed fiber (OZ Optics TSMJ-X-15509/125-0.25-7-2.5-14-2) is mounted on a 3-axis nanopositioner (Attocube ANPx101/LT and ANPz102/LT) and used for photon collection. Photons are detected by a pair of superconducting nanowire single photon detectors (Quantum Opus QO-NPD-1200-1600) with quantum efficiencies of $\sim60\%$. The experimental configuration is shown in Fig~\Ref{figS1}. For taking device reflection spectra and resonantly exciting T centers, an InstaTune FP4209 fast random-access tunable laser is used. We use an Aerodiode semiconductor optical amplifier to generate $200\,\mathrm{ns}$-long pulses with $<1\,\mathrm{ns}$ rise and fall times. Input power is controlled using a variable optical attenuator (Agiltron MSOA-02B1H1333) and measured using a ThorLabs PM100USB power meter. We measure device coupling efficiencies and take device reflection spectra using a high-speed InGaAs photodiode (ThorLabs DET01CFC). We filter signals from our device with an adjustable bandwidth and center wavelength filter (WL Photonics WLTF-BA-U-1310-100-SM-0.9/2.0-FC/APC-USB), and use an acousto-optic modulator (Aerodiode 1310-AOM) to optically gate the detectors to prevent heating from reflected laser pulses. Pulses are generated and synchronized using Swabian Instruments Pulse Streamer 8/2, and time-tagged using Swabian Instruments Time Tagger Ultra.

To introduce nitrogen gas into the cryostat for cavity tuning, we connect a nitrogen source to a needle valve (Edwards LV10K) with PTFE tubing. A small reservoir volume 
between the needle valve and an electroMAG solenoid valve (Ideal Vacuum) is pumped by an Edwards XDD1 backing pump. By adjusting the leak rate of the needle valve, the quantity of nitrogen introduced upon opening the solenoid valve is controlled. We find that a nitrogen pressure of $\sim3$ Torr enables the non-disruptive introduction of nitrogen into the chamber. The nitrogen is introduced into the cryostat through a gas tube side panel assembly from Montana Instruments. The gas tube in the chamber is a 1/16" OD 
stainless steel tube, which is thermally shorted to the $4\,\mathrm{K}$ stage of the cryostat, causing the nitrogen to freeze and clog the tube before reaching the sample. A $50\,\Omega$ resistive heater 
at the end of the tube enables local heating, sublimating a discrete quantity of nitrogen, which condenses onto the sample's surface.

To resonantly tune cavities, we apply $0.1\,\mathrm{s}$ resonant laser pulses at $\sim1\,\mathrm{\mu W}$ peak powers in the bus waveguide, which induces discrete shifts in target cavity resonances on the order of the cavity linewidth ($\sim\,\mathrm{GHz}$). The process is stabilized by negative feedback, where the blue-shift increases laser-cavity detuning, reduces the intra-cavity power, and stops further tuning. We do not observe any crosstalk with off-resonant cavities due to high frequency-selectivity for operation close to the power threshold. We continue applying laser pulses while tracking the magnitude of resonance shifts by fitting the reflection spectrum. By adjusting the laser power between pulses, we maintain a desired tuning step size, typically around $\mathrm{0.5 \,GHz/\mathrm{step}}$. While this method is favored for controlled tuning rates, faster tuning is possible by increasing the laser power to $\sim$ mW and sweeping the laser frequency at rates of $\sim 10\,\mathrm{GHz/s}$.

This method fails when cavities are not spectrally resolvable or when two cavities overlap spectrally. In these cases, we perform the same feedback-based iterative tuning method using a 635nm continuous-wave laser (ThorLabs S1FC635). The laser is focused onto the sample surface from free space through a microscope objective (Mitutoyo LCD Plan Apo NIR 50, NA=0.42), where a variable optical attenuator controls the laser power.
\begin{figure}[t!]
\centering
\includegraphics[width=\columnwidth]{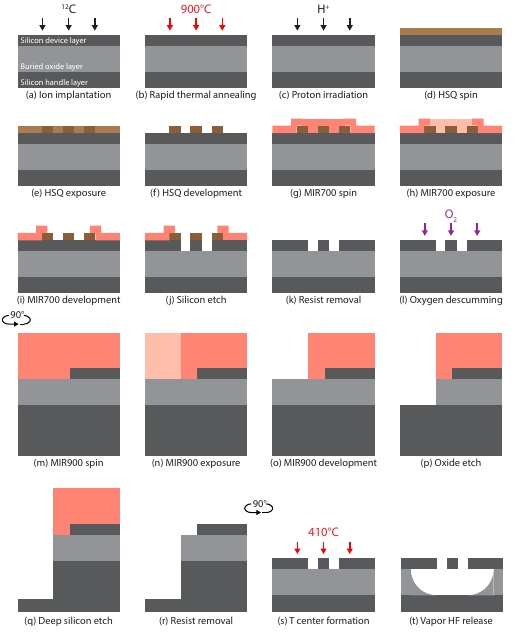}
\caption{\textbf{Device fabrication steps.} (a-l) Front views of the electron-beam and optical lithography steps to create T centers in photonic devices. (m-r) Side views of the facet etch for fiber coupling. (s-t) Front views of the released device containing T centers formed by annealing.}
\label{figS2}
\end{figure}

\section{Device fabrication and design}\label{fab-and-design}

The device is fabricated on a high-resistivity silicon-on-insulator (SOI) substrate from SEH America. The 220 nm-thick device layer has $\geq 3\,\mathrm{k\Omega \,cm}$ resistivity, 100 orientation, and is grown by Float-zone. To create T centers in the device, we perform $^{12}\mathrm{C}$ implantation at $36\,\mathrm{keV}$ with a fluence of $10^{13}\,\mathrm{cm^{-2}}$. We perform rapid thermal annealing at $900^\circ\,\mathrm{C}$ for 20 seconds to repair lattice damage from implantation, then implant $\mathrm{H}$ at $9\,\mathrm{keV}$ with a fluence of $7\cdot 10^{12}\,\mathrm{cm^{-2}}$ to introduce hydrogen for T center formation. Finally, we anneal the device at $410^\circ\,\mathrm{C}$ for 3 minutes to form T centers. Device fabrication and T center creation steps are shown in Fig.~\ref{figS2}.

\begin{table*}[t]\centering
\begin{tabular}{ |p{0.5cm}||p{1.75cm}|p{2.25cm}|p{2.25cm}|p{1.25cm}|p{1.0cm}| }
 \hline
  & $\lambda_T\,\mathrm{(nm)}$ & $\gamma_\mathrm{tot} / 2\pi\,\mathrm{(GHz)}$ & $\kappa_\mathrm{tot} / 2\pi\,\mathrm{(GHz)}$ & $\tau\,\mathrm{(ns)}$ & $P$  \\
 \hline
$C_1$ & $1325.749$ & $3.3$ & $14.6$ & $300.5$ & $9.5$ \\
$C_1$ & $1325.755$ & $3.1$ & $16.9$ & $352.3$ & $7.5$  \\
$C_1$ & $1325.878$ & $7.6$ & $5.1$ & $62.1$ & $62.8$  \\
\hline
$C_2$ & $1325.879$ & - & $12.0$ & $135.8$ & $26.4$  \\
\hline
$C_3$ & $1325.706$ & $4.1$ & $9.7$ & $372.5$ & $6.9$ \\
$C_3$ & $1325.758$ & $2.3$ & $9.6$ & $329.8$ & $8.3$  \\
$C_3$ & $1325.767$ & $4.1$ & $16.9$ & $162.9$ & $21.3$ \\
$C_3$ & $1325.770$ & - & $4.5$ & $50.8$ & $77.8$  \\
\hline
$C_4$ & $1325.736$ & $6.2$ & $14.4$ & $318.4$ & $8.8$  \\
$C_4$ & $1325.737$ & $2.9$ & $13.9$ & $231.0$ & $13.7$ \\
$C_4$ & $1325.758$ & $3.7$ & $16.8$ & $361.0$ & $7.2$ \\
\hline
$C_5$ & $1325.770$ & $4.1$ & $17.4$ & $301.8$ & $9.5$ \\
$C_5$ & $1325.777$ & $5.4$ & $4.4$ & $488.0$ & $4.2$  \\
$C_5$ & $1325.779$ & $5.9$ & $20.1$ & $225.8$ & $14.1$ \\
\hline
$C_6$ & $1325.762$ & $3.6$ & $5.5$ & $238.2$ & $13.2$ \\
$C_6$ & $1325.849$ & $5.4$ & $7.0$ & $89.3$ & $42.4$ \\
\hline
$C_7$ & $1325.712$ & $1.6$ & $1.6$ & $153.0$ & $22.9$ \\
\hline
\end{tabular}
\caption{\textbf{Summary of measured T centers in cavities.} Optical linewidths are power-broadened and are an upper bound. }
\label{table}
\end{table*}

The photonic crystal cavities (PCCs) in this device consist of arrays of circular holes in a $400\,\mathrm{nm}$-wide waveguide which are linearly tapered towards the center of the array to form the cavity. The mirror regions of the PCC consist of $N_m$ unit cells, and the hole dimensions are determined by a filling fraction $f$ and lattice constant $a_m$. The defect region of the PCC consists of $N_t$ tapering unit cells, with a minimum lattice constant $a_d = a_m - 50\,\mathrm{nm}$ in the center. The PCCs are coupled to a $350\,\mathrm{nm}$-wide bus waveguide separated from the PCC by $d_\mathrm{gap}$, which determines the cavity $Q_e$. For $\{f,\, a_m,\, N_m,\, N_d \} = \{0.45,\, 315\,\mathrm{nm},\, 15,\, 7\}$, we simulate the PCC in COMSOL to obtain a resonance at $1350\,\mathrm{nm}$ with $Q_i \sim 10^7$ and a mode volume of $V = 0.5(\lambda_0/n)^3$. The lattice constant is swept from $a_m = 250\,\mathrm{nm}$ to $a_m = 350\,\mathrm{nm}$ in $2\,\mathrm{nm}$ steps in the device to account for fabrication error and ensure overlap with T center wavelengths. We choose $d_\mathrm{gap} = 336\,\mathrm{nm}$ to target $Q_e = 10^5$. We note that measured $Q_i$ and $Q_e$ deviate significantly from the simulated values. 
The discrepancy in $Q_i$ can be explained by loss induced by implantation damage or fabrication errors.

\section{Analysis of Purcell enhancement}\label{purcell-analysis}

The Purcell factor $P$ can be calculated from the cavity-enhanced lifetime of the T center. The spontaneous emission rate is the sum of the radiative ($\gamma_\mathrm{r}$) and non-radiative ($\gamma_\mathrm{nr}$) decay rates, where the radiative decay rate is the sum of the decay rates into the zero-phonon line ($\gamma_\mathrm{ZPL}$) and phonon sideband ($\gamma_\mathrm{PSB}$). The total decay rate $\gamma$ can then be expressed as $\gamma =  \gamma_\mathrm{ZPL} / (\eta_\mathrm{QE} \,\eta_\mathrm{DW})$, where $\eta_\mathrm{QE} = \gamma_{r} / (\gamma_\mathrm{r} + \gamma_{nr})$ is the quantum efficiency and $\eta_\mathrm{DW} = \gamma_{ZPL} / (\gamma_\mathrm{ZPL} + \gamma_{PSB})$ is the Debye-Waller factor. The cavity enhances the decay rate into the zero-phonon line, and the enhanced decay rate is $\gamma^\prime = (1 + P)\gamma_\mathrm{ZPL} + \gamma_\mathrm{PSB} + \gamma_\mathrm{nr} = \gamma + P\,\gamma_\mathrm{ZPL}$. The Purcell factor is then $P = (\gamma^\prime/\gamma - 1) / (\eta_\mathrm{QE} \,\eta_\mathrm{DW})$. The T centers in Fig.~\ref{fig3} have lifetimes of $\tau_{T1} = 213.89(17)\,\mathrm{ns}$ and $\tau_{T2} = 84.32(12)\,\mathrm{ns}$, corresponding lower-bounded Purcell factors of $P_{T1} = 15.16$ and $P_{T2} = 45.15$ assuming perfect quantum efficiencies and a Debye-Waller factor of $0.23$.

\section{Statistics}\label{maps}

We measure a total of 17 T centers across 7 cavities coupled to a single bus waveguide over multiple cool-down and warm-up cycles. The results of these measurements are shown in Table~\ref{table}. The minimum and maximum T center wavelengths are $1325.7056\,\mathrm{nm}$ and $1325.8791\,\mathrm{nm}$, corresponding to a $30\,\mathrm{GHz}$ range. The average Purcell-enhanced lifetime is $240\,\mathrm{ns}$, with the shortest measured lifetime being $50.8\,\mathrm{ns}$. All measured T center linewidths are power broadened, with a minimum measured linewidth of $1.64\,\mathrm{GHz}$. We observe T center wavelengths shifting after warming up the cryostat, implying that some of the 17 measured T centers may be the same center measured during different thermal cycles. A likely cause of the shifts is changes in local strain during thermal cycles.

\section{Modeling the cascaded cavity-emitter array with SLH}\label{SLH-model}

\begin{figure}[hbt!]
    \centering
    \includegraphics[scale=0.3]{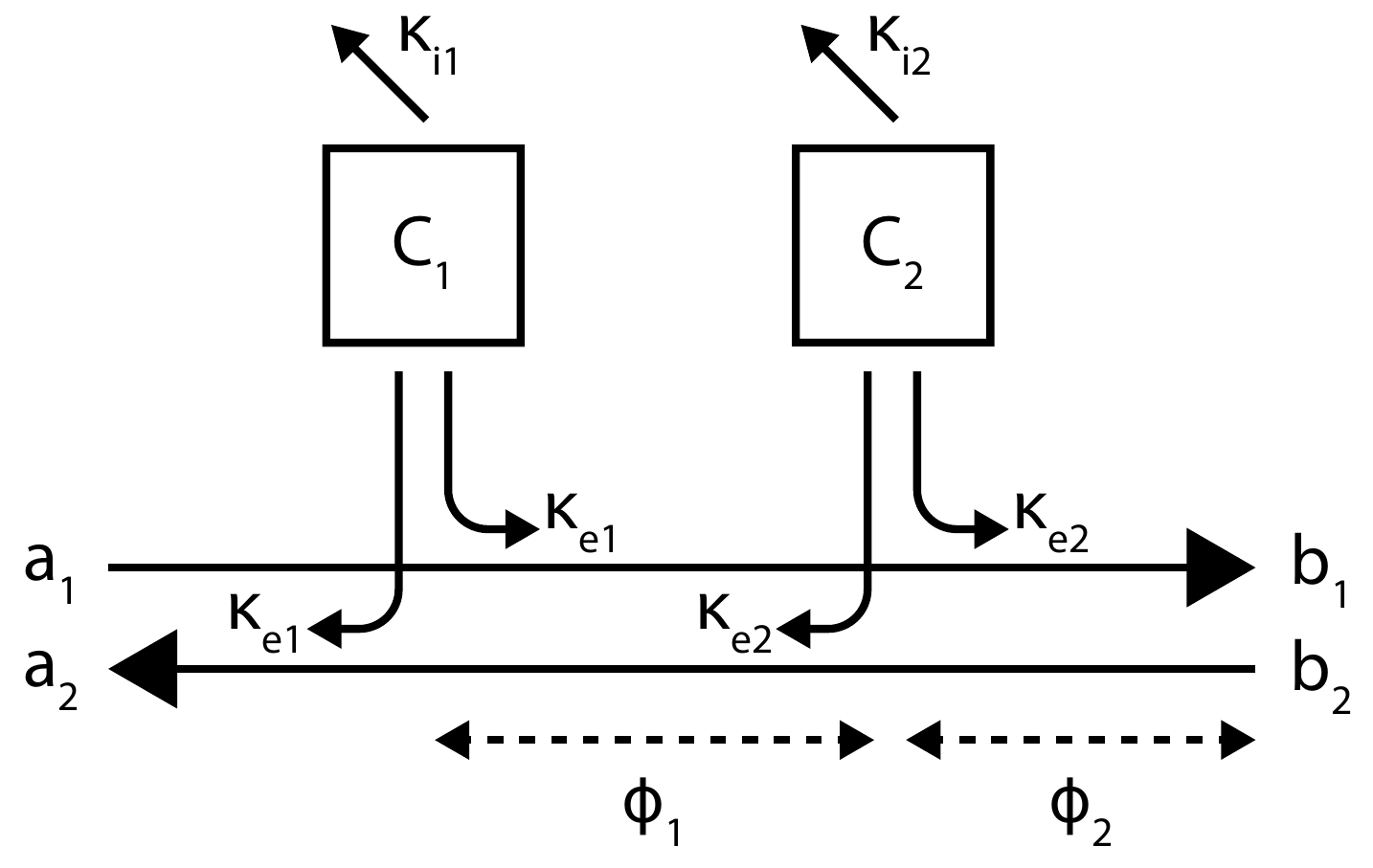}
    \caption{Illustration of the model}
    \label{figSLH1}
\end{figure}

We model our system using the SLH framework~\cite{Combes2017}, a modular framework for modeling networked quantum systems. We choose this framework to account for coherent feedback channels introduced by the terminating mirror in our device. We model our system as two optical cavities ($C_1, C_2$) with resonances detuned by $\Delta_1$, $\Delta_2$ from a probe drive $\omega_p = 2\pi c / \lambda_p$. The cavities are side-coupled to left- and right-propagating modes at rates $\kappa_{e1,e2}$, with intrinsic decay rates $\kappa_{i1,i2}$. The cavities are separated by a distance $L_1$, corresponding to a relative phase of $\phi_1$ = $2\pi \cdot L_1 / \lambda_p$ from the phase accumulated during propagation. We define the input to the right (left) propagating mode as $a_1$ ($b_2$) and the output as $b_1$ ($a_2$). We will capture the effect of the terminating mirror by connecting $b_1$ to $b_2$ at a distance $L_2$ from $C_2$, or at a phase length $\phi_2$ = $2\pi \cdot L_2 / \lambda_p$. In the main text, $\theta_1 = \phi_1 + \phi_2$ and $\theta_2 = \phi_2$ are the phase lengths between the cavities and the mirror. An illustration of our model is shown in Fig~\Ref{figSLH1}.

\begin{figure}[b!]
    \centering
    \includegraphics[scale=0.3]{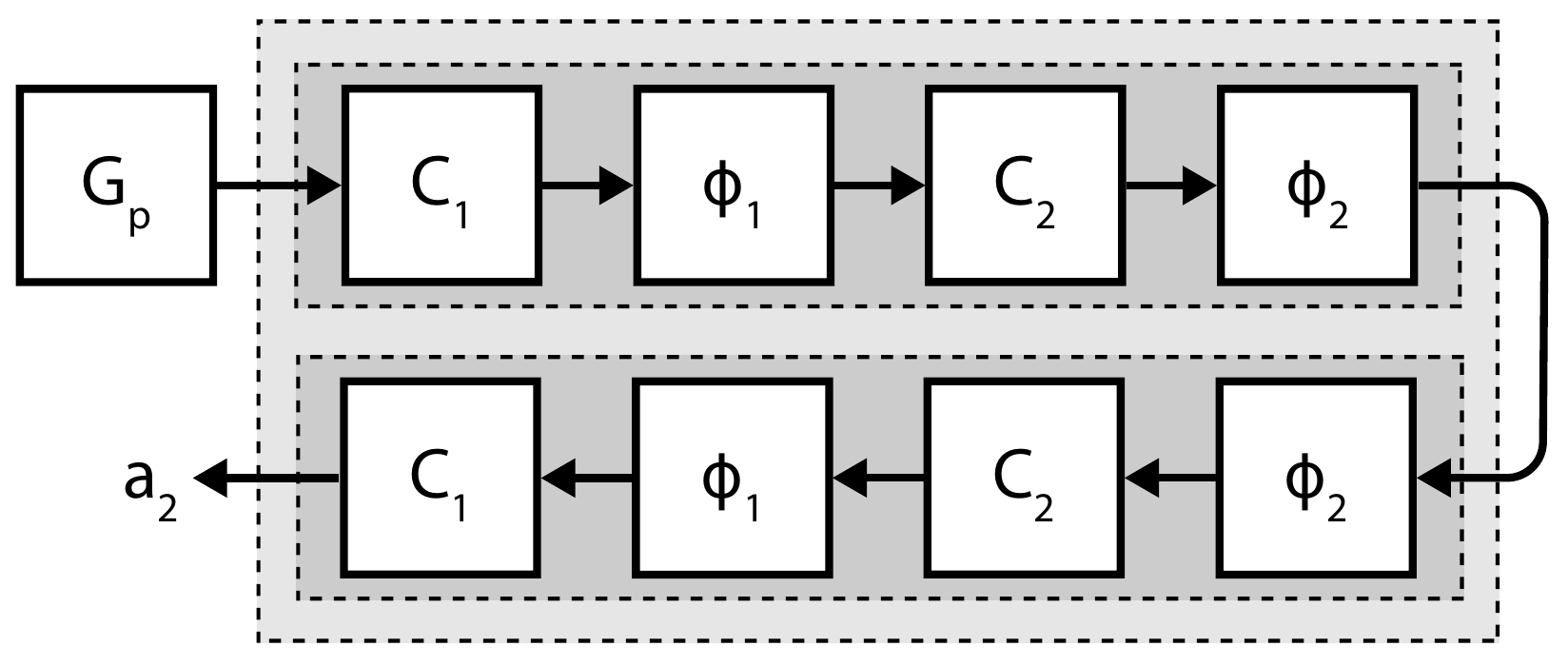}
    \caption{Illustration of the system SLH triple.}
    \label{figSLH2}
\end{figure}

We now define the SLH triples for the components in our system. The SLH triple for a phase shift $\phi$ is
\begin{align}
    G_\phi = \left(e^{i\phi},\, 0,\, 0\right)
\end{align}
The triple for a cavity $C_i$ coupled to the right-propagating mode is
\begin{align}
    G_{c_{i},r} = \left(1,\, \sqrt{\kappa_{ei}/2}\,\hat{a}_i,\, \Delta_i \hat{a}_i^\dagger \hat{a}_i\right)
\end{align}
which differs from the triple for the left-propagating mode
\begin{align}
    G_{c_{i},l} = \left(1,\, \sqrt{\kappa_{ei}/2}\,\hat{a}_i,\,  0 \right)
\end{align}
so as not to double-count the cavity Hamiltonians from the two modes to which they are coupled. 
By cascading the triples in the left- and right-propagating modes, concatenating the two modes, and applying feedback, we obtain the system SLH triple
\begin{equation}\begin{aligned}
    G_\mathrm{sys} &=\Big[\left(G_{\phi_{2}} \lhd G_{c_{2},r} \lhd G_{\phi_{1}} \lhd G_{c_{1},r}\right) \\&\boxplus \left(G_{c_{1},l} \lhd G_{\phi_{1}} \lhd G_{c_{2},l} \lhd G_{\phi_{2}} \right) \Big]_{1\to 2} \lhd G_p
\end{aligned}\end{equation}
where we have added a coherent probe $G_p = \left(1, \alpha, 0\right)$. The system SLH triple is illustrated in Fig~\Ref{figSLH2}.

We use Mathematica to apply the SLH composition rules and calculate $G_\mathrm{sys}$

\begin{equation}\begin{aligned}
    G_\mathrm{sys} = \Big(  &e^{2i(\phi_1+\phi_2)}, \\
                            &e^{i(\phi_1+\phi_2)} \Big( e^{i(\phi_1+\phi_2)}\alpha \\&+ \sqrt{2 \kappa_{e1}} \cos(\phi_1 + \phi_2) \hat{a}_1 + \sqrt{2 \kappa_{e2}} \cos(\phi_2) \hat{a}_2 \Big), \\
                            &\left( \Delta_1 + \frac{\kappa_{e1}}{2} \sin(2\phi_1+2\phi_2) \right)\hat{a}_1^\dagger \hat{a}_1 \\&+ \left( \Delta_2 + \frac{\kappa_{e2}}{2} \sin(2\phi_2) \right)\hat{a}_2^\dagger \hat{a}_2 \\
                            &+ \left( \sqrt{\kappa_{e1}\kappa_{e2}} \cos(\phi_2) \sin(\phi_1+\phi_2) \right) \left( \hat{a}_1^\dagger \hat{a}_2 + \hat{a}_2^\dagger \hat{a}_1 \right)  \\
                            &+ \frac{1}{2i}\sqrt{\frac{\kappa_{e1}}{2}} \left(1 +  e^{2i(\phi_1+\phi_2)} \right) \hat{a}_1^\dagger \alpha \\&+ \frac{1}{2i}\sqrt{\frac{\kappa_{e2}}{2}} \left(e^{i\phi_1} + e^{i(\phi_1+2\phi_2)} \right) \hat{a}_2^\dagger \alpha + \mathrm{h.c.} \Big)
\end{aligned}\end{equation}
We can now write the master equation
\begin{align}
    \dot{\rho} &= -i \left[H, \rho\right] + \mathcal{L}(L)\rho
\end{align}
where the Liouvillian operator is
\begin{equation}\begin{aligned}
    \mathcal{L}(L)\rho &= L\rho L^\dagger - \frac{1}{2}\left(L^\dagger L \rho + \rho L^\dagger L \right) \\
    &= 2\kappa_{e1}\cos^2(\phi_1+\phi_2) \left( \hat{a}_1\rho \hat{a}_1^\dagger - \frac{1}{2}\hat{a}_1^\dagger \hat{a}_1 \rho - \frac{1}{2}\rho \hat{a}_1^\dagger \hat{a}_1 \right) \\
    &+ 2\kappa_{e2}\cos^2(\phi_2) \left( \hat{a}_2\rho \hat{a}_2^\dagger - \frac{1}{2}\hat{a}_2^\dagger \hat{a}_2 \rho - \frac{1}{2}\rho \hat{a}_2^\dagger \hat{a}_2 \right) \\
    &+ 2 \sqrt{\kappa_{e1}\kappa_{e2}} \cos(\phi_2)\cos(\phi_1+\phi_2) \Big( \hat{a}_1\rho \hat{a}_2^\dagger - \frac{1}{2}\hat{a}_1^\dagger \hat{a}_2 \rho \\&- \frac{1}{2}\rho \hat{a}_1^\dagger \hat{a}_2 + \hat{a}_2\rho \hat{a}_1^\dagger - \frac{1}{2}\hat{a}_2^\dagger \hat{a}_1 \rho - \frac{1}{2}\rho \hat{a}_2^\dagger \hat{a}_1 \Big) \\
    &+ \frac{1}{2}\sqrt{\frac{\kappa_{e1}}{2}} \left(1 +  e^{2i(\phi_1+\phi_2)} \right) \hat{a}_1^\dagger \alpha \rho \\&+ \frac{1}{2}\sqrt{\frac{\kappa_{e2}}{2}} \left(e^{i\phi_1} + e^{i(\phi_1+2\phi_2)} \right) \hat{a}_2^\dagger \alpha \rho + \mathrm{h.c.}
\end{aligned}\end{equation}
We now make the following substitutions
\begin{equation}\begin{aligned}
    \widetilde{\Delta}_1 &= \Delta_1 + \frac{\kappa_{e1}}{2} \sin(2\theta_1)\\
    \widetilde{\Delta}_2 &=  \Delta_2 + \frac{\kappa_{e2}}{2} \sin(2\theta_2) \\
    g_c &= \frac{\sqrt{\kappa_{e1}\kappa_{e2}}}{2} \left( \sin(\theta_1 + \theta_2) + \sin(\theta_1 - \theta_2) \right) \\
    \Omega_1 &= \frac{\alpha}{i}\sqrt{\frac{\kappa_{e1}}{2}} \left(1 +  e^{2i\theta_1} \right) \\
    \Omega_2 &= \frac{\alpha}{i}\sqrt{\frac{\kappa_{e2}}{2}} \left(e^{i(\theta_1 + \theta_2)} + e^{i(\theta_1 - \theta_2)} \right) 
\end{aligned}\end{equation}
where we have reintroduced the cavity-mirror phase lengths $\theta_1 = \phi_1 + \phi_2$ and $\theta_2 = \phi_2$. These substitutions give rise to $\theta_1 \pm \theta_2$ terms originating from the two interference paths in the system. We can now rewrite the Hamiltonian as
\begin{equation}\begin{aligned}
    \hat{H} &= \widetilde{\Delta}_1 \hat{a}_1^\dagger \hat{a}_1 + \widetilde{\Delta}_2 \hat{a}_2^\dagger \hat{a}_2 + g_c \left( \hat{a}_1^\dagger \hat{a}_2 + \hat{a}_2^\dagger \hat{a}_1 \right) \\&+ \frac{1}{2}\left(\Omega_1 \hat{a}_1^\dagger + \Omega_2 \hat{a}_2^\dagger + \mathrm{h.c.}\right)\Big)
\end{aligned}\end{equation}
We make similar substitutions for $\mathcal{L}(L)\rho$
\begin{equation}\begin{aligned}
    \widetilde{\kappa}_{1} &= 2\kappa_{e1}\cos^2(\theta_1) + \kappa_{i1} \\
    \widetilde{\kappa}_{2} &= 2\kappa_{e2}\cos^2(\theta_2) + \kappa_{i2} \\
    {\kappa}_{c} &= \sqrt{\kappa_{e1}\kappa_{e2}}\left( \cos(\theta_1 + \theta_2) + \cos(\theta_1 - \theta_2) \right)
\end{aligned}\end{equation}
where we have added the intrinsic cavity decay rates $\kappa_{i1,i2}$. We can now rewrite $\mathcal{L}(L)\rho$ 
\begin{equation}\begin{aligned}
    \mathcal{L}(L)\rho &= \widetilde{\kappa}_{1} \left( \hat{a}_1\rho \hat{a}_1^\dagger - \frac{1}{2}\hat{a}_1^\dagger \hat{a}_1 \rho - \frac{1}{2}\rho \hat{a}_1^\dagger \hat{a}_1 \right) \\
    &+ \widetilde{\kappa}_{2} \left( \hat{a}_2\rho \hat{a}_2^\dagger - \frac{1}{2}\hat{a}_2^\dagger \hat{a}_2 \rho - \frac{1}{2}\rho \hat{a}_2^\dagger \hat{a}_2 \right) \\
    &+ \widetilde{\kappa}_{c} \Big( \hat{a}_1\rho \hat{a}_2^\dagger - \frac{1}{2}\hat{a}_1^\dagger \hat{a}_2 \rho - \frac{1}{2}\rho \hat{a}_1^\dagger \hat{a}_2 \\&+ \hat{a}_2\rho \hat{a}_1^\dagger - \frac{1}{2}\hat{a}_2^\dagger \hat{a}_1 \rho - \frac{1}{2}\rho \hat{a}_2^\dagger \hat{a}_1 \Big) \\
    &+ \frac{1}{2} i \Omega_1 \hat{a}_1^\dagger \rho + \frac{1}{2} i \Omega_2 \hat{a}_2^\dagger \rho + \mathrm{h.c.}
\end{aligned}\end{equation}
Allowing us to write the complete master equation
\begin{equation}\begin{aligned}
    \dot{\rho} = &-i \Big[\widetilde{\Delta}_1 \hat{a}_1^\dagger \hat{a}_1 + \widetilde{\Delta}_2 \hat{a}_2^\dagger \hat{a}_2 + g_c \left( \hat{a}_1^\dagger \hat{a}_2 + \hat{a}_2^\dagger \hat{a}_1 \right) \\&+ \Omega_1 \hat{a}_1^\dagger + \Omega_2 \hat{a}_2^\dagger + \mathrm{h.c.}\,, \rho\Big] \\
    &+ \widetilde{\kappa}_{1} \left( \hat{a}_1\rho \hat{a}_1^\dagger - \frac{1}{2}\hat{a}_1^\dagger \hat{a}_1 \rho - \frac{1}{2}\rho \hat{a}_1^\dagger \hat{a}_1 \right) \\
    &+ \widetilde{\kappa}_{2} \left( \hat{a}_2\rho \hat{a}_2^\dagger - \frac{1}{2}\hat{a}_2^\dagger \hat{a}_2 \rho - \frac{1}{2}\rho \hat{a}_2^\dagger \hat{a}_2 \right) \\
    &+ \widetilde{\kappa}_{c} \Big( \hat{a}_1\rho \hat{a}_2^\dagger - \frac{1}{2}\hat{a}_1^\dagger \hat{a}_2 \rho - \frac{1}{2}\rho \hat{a}_1^\dagger \hat{a}_2 \\&+ \hat{a}_2\rho \hat{a}_1^\dagger - \frac{1}{2}\hat{a}_2^\dagger \hat{a}_1 \rho - \frac{1}{2}\rho \hat{a}_2^\dagger \hat{a}_1 \Big) 
\end{aligned}\end{equation}
We can find the expectation value of the cavity operators
\begin{equation}\begin{aligned}
    \langle \hat{a}_1 \rangle &= \Tr( \hat{a}_1 \rho ) \\
    \frac{d \langle \hat{a}_1 \rangle}{dt} &= \Tr( \hat{a}_1 \frac{d\rho}{dt} ) \\
    &= -\left(i\widetilde{\Delta}_1 + \frac{1}{2} \widetilde{\kappa}_{1} \right) \langle \hat{a}_1 \rangle - \left( i g_c + \frac{1}{2} \widetilde{\kappa}_{c}  \right) \langle \hat{a}_2 \rangle - i\Omega_1 \\
    \frac{d \langle \hat{a}_2 \rangle}{dt} &= -\left(i\widetilde{\Delta}_2 + \frac{1}{2} \widetilde{\kappa}_{2} \right) \langle \hat{a}_2 \rangle - \left( i g_c + \frac{1}{2} \widetilde{\kappa}_{c}  \right) \langle \hat{a}_1 \rangle - i\Omega_2 
\end{aligned}\end{equation}
We find the steady-state solutions ($\frac{d \langle \hat{a}_1 \rangle}{dt} = \frac{d \langle \hat{a}_2 \rangle}{dt} = 0$)
\begin{equation}\begin{aligned}
    \langle \hat{a}_1 \rangle &= \frac{i\Omega_2 \left( i g_c + \frac{1}{2} \widetilde{\kappa}_{c}  \right) - i\Omega_1 \left(i\widetilde{\Delta}_2 + \frac{1}{2} \widetilde{\kappa}_{2} \right)}{\left( i g_c + \frac{1}{2} \widetilde{\kappa}_{c}  \right)^2 - \left(i\widetilde{\Delta}_1 + \frac{1}{2} \widetilde{\kappa}_{1} \right)\left(i\widetilde{\Delta}_2 + \frac{1}{2} \widetilde{\kappa}_{2} \right)} \\
    \langle \hat{a}_2 \rangle &= \frac{ i\Omega_2 \left(i\widetilde{\Delta}_1 + \frac{1}{2} \widetilde{\kappa}_{1} \right) - i\Omega_1 \left( i g_c + \frac{1}{2} \widetilde{\kappa}_{c} \right) }{\left( i g_c + \frac{1}{2} \widetilde{\kappa}_{c}  \right)^2 - \left(i\widetilde{\Delta}_1 + \frac{1}{2} \widetilde{\kappa}_{1} \right)\left(i\widetilde{\Delta}_2 + \frac{1}{2} \widetilde{\kappa}_{2} \right)}
\end{aligned}\end{equation}
From which we calculate the reflection spectrum
\begin{equation}\begin{aligned}
    L/\alpha &= e^{i\theta_1} \Big( e^{i\theta_1}\alpha \\&+ \sqrt{2 \kappa_{e1}} \cos(\theta_1) \langle \hat{a}_1 \rangle + \sqrt{2 \kappa_{e2}} \cos(\theta_2) \langle \hat{a}_2 \rangle \Big)
    \label{model-ref-spec}
\end{aligned}\end{equation}

To model the emitter's lifetime during cavity hybridization, we write the effective Hamiltonian
\begin{equation}\begin{aligned}
    \hat{H}_\mathrm{eff} &= \left(\widetilde{\Delta}_1 - i\frac{\widetilde{\kappa}_{1}}{2}\right)\hat{a}_1^\dagger \hat{a}_1 + \left( \widetilde{\Delta}_2 - i\frac{\widetilde{\kappa}_{2}}{2} \right) \hat{a}_2^\dagger \hat{a}_2 \\&+ \left( g_c - i\frac{\widetilde{\kappa}_{c}}{2} \right) \left( \hat{a}_1^\dagger \hat{a}_2 + \hat{a}_2^\dagger \hat{a}_1 \right)
    \label{h-eff-no-emitter}
\end{aligned}\end{equation}
where we have neglected drive terms. To include the emitter in our model, we introduce emitter terms as a weak perturbation
\begin{equation}\begin{aligned}
    \hat{H}_\mathrm{eff} = &\left(\widetilde{\Delta}_1 - i\frac{\widetilde{\kappa}_{1}}{2}\right)\hat{a}_1^\dagger \hat{a}_1 + \left( \widetilde{\Delta}_2 - i\frac{\widetilde{\kappa}_{2}}{2} \right) \hat{a}_2^\dagger \hat{a}_2 \\&+ \left( g_c - i\frac{\widetilde{\kappa}_{c}}{2} \right) \left( \hat{a}_1^\dagger \hat{a}_2 + \hat{a}_2^\dagger \hat{a}_1 \right) \\
    &+ g_e \left( \sigma_+ \hat{a}_1 + \hat{a}_1^\dagger \sigma_- \right) + \left(\Delta_e - i\frac{\kappa_q}{2}\right) \sigma_+ \sigma_-
    \label{h-eff-with-emitter}
\end{aligned}\end{equation}
where $\Delta_e$ and $\kappa_q$ are the emitter detuning and decay rate. We diagonalize the Hamiltonian in the single-excitation manifold 
to obtain

\begin{equation}\begin{aligned}
    \hat{H}_\mathrm{diag} = &\left({\Delta}_{b_1} - i\frac{{\kappa}_{b_1}}{2}\right)\hat{b}_1^\dagger \hat{b}_1 + \left({\Delta}_{b_2} - i\frac{{\kappa}_{b_2}}{2}\right)\hat{b}_2^\dagger \hat{b}_2 \\&+ \left(\widetilde{\Delta}_e - i\frac{\widetilde{\kappa}_q}{2}\right) \hat{b}_3^\dagger \hat{b}_3 
    \label{h-eff-diag}
\end{aligned}\end{equation}
where $b_1$ and $b_2$ correspond to the hybridized cavity modes and $b_3$ corresponds to the weakly hybridized emitter mode. $\Delta_{b_{1,2}}$ and $\kappa_{b_{1,2}}$ correspond to the hybrid mode detunings and decay rates, $\widetilde{\Delta}_e$ is the detuning of the hybridized emitter, and $\widetilde{\kappa}_q$ is the modified decay rate of the emitter.

From the decomposition of the eigenstates
\begin{equation}\begin{aligned}
    \hat{b}_1 &= \alpha_1 \hat{a}_1 + \beta_1 \hat{a}_2 + \gamma_1 \sigma_- \\
    \hat{b}_2 &= \alpha_2 \hat{a}_1 + \beta_2 \hat{a}_2 + \gamma_2 \sigma_- \\
    \hat{b}_3 &= \alpha_3 \hat{a}_1 + \beta_3 \hat{a}_2 + \gamma_3 \sigma_-
\end{aligned}\end{equation}

We can obtain the amplitude of the hybrid modes in terms of the bare cavity modes ($|\alpha_1|^2$ in $C_1$ for hybrid mode $b_1$) and the phases ($\arctan(\Im \alpha_1 / \Re \alpha_1)$ in $C_1$ for hybrid mode $b_1$) from the eigenstates.

\begin{figure}[t!]
\centering
\includegraphics[width=\columnwidth]{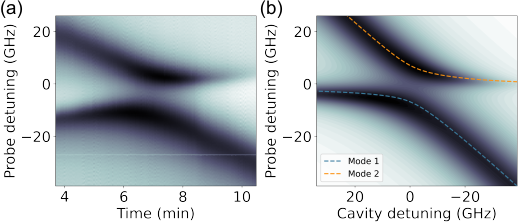}
\caption{\textbf{Strong cavity-cavity coupling.} 
(a) Device reflection spectra during two-cavity hybridization
showing an avoided crossing.
(b) Corresponding model fit to (a).}
\label{figSN}
\end{figure}

To fit the data from Fig.~\ref{fig4} to our model, we first fit the reflection spectrum (Eq.~\ref{model-ref-spec}) to Fig.~\ref{fig4}(b), obtaining the set of parameters $\{\lambda_{c1}=1325.9132\,\mathrm{nm},\, Q_e=10165,\, Q_{i1}=35460,\, Q_{i2}=34441,\, \phi_1=0.78\pi,\, \phi_2=1.44\pi\}$, and an array of values for $\lambda_{c2}$. We constrain all model parameters to be constant during cavity tuning except $\lambda_{c2}$, and set $Q_{e1} = Q_{e2} = Q_{e}$. Using these parameters, we diagonalize the effective Hamiltonian (Eq.~\ref{h-eff-with-emitter}) for every $\lambda_{c2}$, obtaining the parameters $\{\alpha_{1,2,3},\, \beta_{1,2,3},\, \gamma_{1,2,3}\}$. From these values, we calculate the decay rates of the hybrid modes in Fig.~\ref{fig4}(c), the populations of the modes in $C_1$ for Fig.~\ref{fig4}(d), and the lifetimes in Fig.~\ref{fig4}(e).

\section{Strong coupling in waveguide-QED}\label{SLH-strong}

A unique feature of our mirror-terminated design is the enhanced cavity-cavity coupling compared to the no-mirror case. This enables the observation of signatures of strong coupling between cavities, such as avoided crossings.

We first consider the case of two cavities $l,m$ separated by a phase $\phi = \theta_l - \theta_m$ and coupled symmetrically to a bus waveguide at a rate $\kappa_e$. We neglect the intrinsic decay $\kappa_i$ in this analysis. The effective Hamiltonian for this system is
\begin{equation}\begin{aligned}
    \hat{H}_\mathrm{eff}^{(lm)} = &\left(\Delta_l - i\frac{{\kappa_e}}{2}\right)\hat{a}_l^\dagger \hat{a}_l + \left( {\Delta}_m - i\frac{{\kappa_e}}{2} \right) \hat{a}_m^\dagger \hat{a}_m \\&+ \left( g_{lm} - i\frac{{\kappa}_{c}}{2} \right) \left( \hat{a}_l^\dagger \hat{a}_m + \hat{a}_m^\dagger \hat{a}_l \right)
\end{aligned}\end{equation}
where $g_{lm} = (\kappa_e/2)\sin\phi$ and $\kappa_{c} = \kappa_e \cos\phi$. When $\Delta_l = \Delta_m = 0$, the effective Hamiltonian can be diagonalized to the following form
\begin{equation}\begin{aligned}
    \hat{H}_\mathrm{eff}^{(lm)} &= \left(g_{lm} - \frac{{i}}{2}\left(\kappa_e + \kappa_c\right)\right)\hat{a}_l^\dagger \hat{a}_l \\&+ \left(-g_{lm} - \frac{{i}}{2}\left(\kappa_e - \kappa_c\right) \right) \hat{a}_m^\dagger \hat{a}_m
\end{aligned}\end{equation}
from which we can extract the hybrid decay rates $\kappa_{h_1} = \kappa_e + \kappa_c$ and $\kappa_{h_2} = \kappa_e - \kappa_c$. We can now calculate the cavity-cavity cooperativity $4g_{lm}^2 / (\kappa_{h_1} \kappa_{h_2})$
\begin{equation}\begin{aligned}
    \frac{4g_{lm}^2}{\kappa_{h_1} \kappa_{h_2}} &= \frac{\kappa_e^2 \sin^2\phi}{(\kappa_e + \kappa_e \cos\phi) (\kappa_e - \kappa_e \cos\phi)} = 1
\end{aligned}\end{equation}
In the general case, the limit $4g_{lm}^2 / (\kappa_{h_1} \kappa_{h_2}) \leq 1$ holds. However, with the addition of a terminating mirror, the cavity-cavity cooperativity is unbounded in general. This distinction from the no-mirror case is due to the mirror enabling conditions where the $g_{lm}$ remains finite while $\kappa_{h_1}$ or $\kappa_{h_2}$ go to zero. We can see this by performing the same calculation with a terminating mirror. In this case, the effective Hamiltonian at zero detuning is

\begin{equation}\begin{aligned}
    \hat{H}_\mathrm{eff}^{(lm)} &= \left(- i\frac{\widetilde{\kappa}_{l}}{2}\right)\hat{a}_l^\dagger \hat{a}_l + \left( - i\frac{\widetilde{\kappa}_{m}}{2} \right) \hat{a}_m^\dagger \hat{a}_m \\&+ \left( g_{lm} - i\frac{{\kappa}_{c}}{2} \right) \left( \hat{a}_l^\dagger \hat{a}_m + \hat{a}_m^\dagger \hat{a}_l \right)
\end{aligned}\end{equation}

where 

\begin{equation}\begin{aligned}
    \widetilde{\kappa}_l &= 2\kappa_{e}\cos^2(\theta_l) \\
    \widetilde{\kappa}_m &= 2\kappa_{e}\cos^2(\theta_m) \\
    g_{lm} &= \frac{\kappa_e}{2}\left( \sin(\theta_l + \theta_m) + \sin(\theta_l - \theta_m) \right) \\
    \kappa_{c} &= \kappa_e \left( \cos(\theta_l + \theta_m) + \cos(\theta_l - \theta_m) \right)
\end{aligned}\end{equation}
We can diagonalize the effective Hamiltonian
\begin{equation}\begin{aligned}
    \hat{H}_\mathrm{eff}^{(lm)} &= -\frac{1}{2}\Big( \frac{i}{2}\widetilde{\kappa}_l + \frac{i}{2}\widetilde{\kappa}_m \\
    &+ \sqrt{- \left( \frac{\widetilde{\kappa}_l}{2} - \frac{\widetilde{\kappa}_m}{2} \right)^2 + 4 \left( g_{lm} - \frac{i}{2}\kappa_c \right)^2} \Big) \hat{a}_l^\dagger \hat{a}_l \\
    &-\frac{1}{2}\Big( \frac{i}{2}\widetilde{\kappa}_l + \frac{i}{2}\widetilde{\kappa}_m \\
    &- \sqrt{- \left( \frac{\widetilde{\kappa}_l}{2} - \frac{\widetilde{\kappa}_m}{2} \right)^2 + 4  \left( g_{lm} - \frac{i}{2}\kappa_c \right)^2} \Big) \hat{a}_m^\dagger \hat{a}_m
\end{aligned}\end{equation}
We now examine the case $\theta_l = \pi/2$, $\theta_m = \pi/4$. Under these conditions, $\widetilde{\kappa}_l = 0$ and $\widetilde{\kappa}_m = \kappa_e$, and the interaction terms are $g_{lm} = \kappa_e/\sqrt{2}$ and $\kappa_c=0$. Therefore, the effective Hamiltonian becomes
\begin{equation}\begin{aligned}
    \hat{H}_\mathrm{eff}^{(lm)} &= -\frac{1}{2}\left( \frac{i}{2}{\kappa}_e + \frac{\sqrt{7}}{2} {\kappa}_e \right) \hat{a}_l^\dagger \hat{a}_l - \frac{1}{2}\left( \frac{i}{2}{\kappa}_e - \frac{\sqrt{7}}{2} {\kappa}_e \right) \hat{a}_m^\dagger \hat{a}_m 
\end{aligned}\end{equation}
where the hybrid decay rates are $\kappa_{h_1} = \kappa_{h_2} = \kappa_e/2$, resulting in $4g_{lm}^2 / (\kappa_{h_1} \kappa_{h_2}) = 8$. 
We experimentally observe avoided crossings between cavities. The data and model are shown in Fig~\Ref{figSN}. At maximum hybridization, the decay rates are measured to be $\kappa_{1}/2\pi = 18.6\,\mathrm{GHz}$ and $\kappa_{2}/2\pi = 11.5\,\mathrm{GHz}$. The strength of the coherent interaction is $g_c/2\pi = 8.6\,\mathrm{GHz}$, resulting in a cavity-cavity cooperativity of $4 g_c^2 / (\kappa_1 \kappa_2) = 1.38$.

\end{document}